\documentclass[twocolumn,pre,twoside,floatfix]{revtex4}

\usepackage{amsmath}
\usepackage{amssymb}
\usepackage{graphicx}
\usepackage{dcolumn}
\usepackage{bm}
\usepackage{color}

\def\dps{\displaystyle}

\def\Eq#1{(\ref{eq:#1})}

\def\epsilon{\varepsilon}
\def\theta{\vartheta}
\def\rho{\varrho}

\begin{document}


\title{Density functional theory of electrowetting}

\author{Markus Bier}
\email{bier@is.mpg.de}
\author{Ingrid Ibagon}
\email{ingrid@is.mpg.de}
\affiliation
{
   Max-Planck-Institut f\"ur Intelligente Systeme, 
   Heisenbergstr.\ 3,
   70569 Stuttgart,
   Germany, 
   and
   Institut f\"ur Theoretische Physik IV,
   Universit\"at Stuttgart,
   Pfaffenwaldring 57,
   70569 Stuttgart,
   Germany
}

\date{10 February 2014}

\begin{abstract}
The phenomenon of electrowetting, i.e., the dependence of the macroscopic
contact angle of a fluid on the electrostatic potential of the substrate, 
is analyzed in terms of the density functional theory of wetting.
It is shown that electrowetting is not an electrocapillarity effect, i.e.,
it cannot be consistently understood in terms of the variation of the
substrate-fluid interfacial tension with the electrostatic substrate
potential, but it is related to the depth of the effective interface potential.
The key feature, which has been overlooked so far and which occurs 
naturally in the density functional approach is the structural change
of a fluid if it is brought into contact with another fluid.
These structural changes occur in the present context as the formation of finite
films of one fluid phase in between the substrate and the bulk of the other fluid
phase.
The non-vanishing Donnan potentials (Galvani potential differences) across such
film-bulk fluid interfaces, which generically occur due to an unequal partitioning
of ions as a result of differences of solubility contrasts, lead to correction terms
in the electrowetting equation, which become relevant for sufficiently small
substrate potentials. 
Whereas the present density functional approach confirms the commonly used
electrocapillarity-based electrowetting equation as a
good approximation for the cases of metallic electrodes or electrodes coated
with a hydrophobic dielectric in contact with an electrolyte solution and an
ion-free oil, a significantly reduced tendency for electrowetting is predicted
for electrodes coated with a dielectric which is hydrophilic or which is in
contact with two immiscible electrolyte solutions.
\end{abstract}

\maketitle


\section{Introduction}

Since the pioneering work of Lippmann \cite{Lippmann1875} and Pellat 
\cite{Pellat1894, Pellat1895} on the influence of electrostatic potentials on 
the wetting of substrates by fluids, electrowetting has been simultaneously 
studied to address fundamental issues of surface science, e.g., 
electrocapillarity \cite{Grahame1947}, the structure of solid-fluid 
interfaces \cite{Sparnaay1964}, or the characterization of surface states 
\cite{Holly1977}, as well as to develop novel applications, e.g., driving,
mixing, or shaping of droplets in lab-on-a-chip devices, optical applications,
or microelectromechanical systems \cite{Mugele2005}.
In the past electrowetting at low voltages was commonly interpreted as an 
electrocapillarity effect, i.e., it is assumed to hinge on the 
voltage-dependence of the substrate-fluid interfacial tension 
\cite{Pellat1894, Pellat1895, Sparnaay1964, Sondag-Huethorst1992, Berge1993,
Vallet1996, Welters1998, Blake2000, Decamps2000, Quinn2003, Jones2004,
Mugele2005, Quinn2005, Klarman2011, Sedev2011, Daub2012}.
A justification for this approach is frequently given in terms of the vast
experimental evidence for systems of uncoated and hydrophobically coated 
electrodes.

The present work reports on an effort to understand electrowetting in the 
context of general wetting phenomena \cite{Dietrich1988}.
Within classical microscopic density functional theory one has
access to the interfacial structure of fluids so that one can study, e.g.,
the contact angle as a function of the electrostatic substrate potential.
However, it turned out that the commonly given derivations of the 
electrowetting equation \cite{Mugele2005} are incorrect in that interfacial
properties, e.g., interfacial tensions or differential capacitances, which
describe the contact of \emph{one} fluid with a substrate, enter the equation 
describing the contact of \emph{two} fluids with a substrate. 
It has been overlooked that the interfacial structure, and thus interfacial
quantities, of a fluid can change upon bringing it into contact with another
fluid.
Although the interfacial structure close of fluids to substrates has been deeply
examined in the surface science literature \cite{Hunter1981, Lyklema1991,
Churaev1992, Lyklema1995, Adamson1997, Hunter2001, Churaev2003}, its
properties seem to be largely ignored in the context of electrowetting so far.
By ignoring structural differences which occur at substrate-fluid 
interfaces upon bringing two fluids in simultaneous contact with a substrate, one
can interpret electrowetting as a consequence of voltage-dependent interfacial
tensions, which is referred to as the electrocapillarity approach to 
electrowetting in the following.
It is shown in this work that electrowetting \emph{cannot} be consistently
understood as an electrocapillarity effect.
Alternative approaches to interpret electrowetting as a line tension
effect have been proposed \cite{Digilov2000} but some of the
predictions were in disagreement with experimental data \cite{Quilliet2001}. 

The present approach is to study electrowetting in terms of the the effective
interface potential \cite{Dietrich1988}, which is related to the macroscopic
contact angle.
The effective interface potential has been analyzed recently for simple 
models of electrolyte solutions \cite{Denesyuk2003, Denesyuk2004, 
Ibagon2013}, showing the general feature of ions inducing wetting transitions
of first order.
Whereas wetting transitions are concerned with the \emph{thickness} of wetting
films, the contact angle is related to the \emph{depth} of the effective
interface potential, which always vanishes continuously at wetting transitions
\cite{Dietrich1988}.
In the following the effective interface potential for Pellat's classical setup
of a vertical parallel plate capacitor \cite{Pellat1894, Pellat1895} is 
determined and used to derive an electrowetting equation 
(Sec.~\ref{sec:theory}) based on the density functional theory of wetting.
This setup has been chosen because its geometry is precisely defined, an
issue which has been recently raised in a critical discussion of the more
common setup of a droplet on a substrate with the counter electrode being a
thin wire \cite{Klarman2011}.
The misconception underlying the classical derivation of the 
electrowetting equation within the electrocapillarity approach is 
discussed (Sec.~\ref{sec:discussion}).
Interestingly, the electrocapillarity-based electrowetting equation
seems to be a good approximation for systems investigated up to now, i.e., 
uncoated or hydrophobically coated electrodes (Sec.~\ref{sec:discussion}).
However, it is argued that the difference in the predicted electrowetting
numbers between the electrocapillarity approach and the present one based on
the density function theory of wetting can be expected to occur, e.g., for 
hydrophilically coated electrodes or for two immiscible electrolyte solutions as
fluids (Sec.~\ref{sec:discussion}).
In view of the conceptional problems of the electrocapillarity approach it is
suggested to rather interpret electrowetting in terms of the density
functional theory of wetting (Sec.~\ref{sec:conclusions}).
Moreover, the possibility to obtain microscopic information about solid-fluid 
interfaces by analyzing electrowetting measurements in terms of the density
functional theory of wetting deserves further consideration. 


\section{\label{sec:theory}Theoretical considerations}

\subsection{Setting}

\begin{figure}[!t]
   \includegraphics{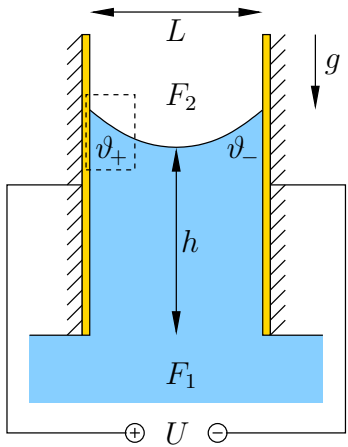}
   \caption{Pellat's setup \cite{Pellat1894, Pellat1895} of electrocapillary
            rise of a fluid $F_1$ in a vertical parallel plate capacitor of width
            $L$ initially filled with a fluid $F_2$.
            The meniscus height $h$ is related to the contact angles $\theta_+$
            and $\theta_-$ by Eq.~\Eq{capillaryrise}.
            Electrowetting corresponds to a dependence of $\theta_+$ and 
            $\theta_-$, and hence of $h$, on the voltage $U$ between the plate
            electrodes.
            A closeup of the three-phase contact region marked by the dashed
            box is depicted in Fig.~\ref{fig:2}.}
   \label{fig:1}
\end{figure}

Consider Pellat's classical setup \cite{Pellat1894, Pellat1895} depicted in 
Fig.~\ref{fig:1}. 
A vertical parallel plate capacitor of width $L$ is in contact with two 
immiscible fluids $F_1$ and $F_2$ of mass densities $\rho_{\text{m}1}$ and
$\rho_{\text{m}2}$, respectively.
At least one of the fluids $F_1$ and $F_2$ is assumed to be an electrolyte solution.
It is further assumed that $\rho_{\text{m}1}>\rho_{\text{m}2}$ so that both
fluids are separated in the gravitational field with $F_1$ being the lower and
$F_2$ being the upper phase (see Fig.~\ref{fig:1}).
Provided the capacitor width $L$ is smaller than the capillary length 
\cite{Rowlinson2002, DeGennes2004, Landau2005}
\begin{align}
   \lambda = \sqrt{\frac{\gamma_{12}}{(\rho_{\text{m}1}-\rho_{\text{m}2})g}}
   \label{eq:capillarylength}
\end{align}
with the $F_1$-$F_2$ interfacial tension $\gamma_{12}$ and the acceleration due
to gravity $g$, the contact angles $\theta_+$ and $\theta_-$ of phase $F_1$ are
related to the meniscus height $h$ by \cite{Rowlinson2002, DeGennes2004, Landau2005}
\begin{align}
   \cos{\theta_+} + \cos\theta_- \simeq \frac{hL}{\lambda^2}
   \quad\text{for $L\ll\lambda$}.
   \label{eq:capillaryrise}
\end{align}
Depending on the interactions of the fluids $F_1, F_2$ and the substrates
$S_+, S_-$, which are metal electrodes (represented by the hatched parts in 
Fig.~\ref{fig:1}) possibly coated with some dielectric (represented by yellow layers
on top of the electrodes in Fig.~\ref{fig:1}), the respective contact angles $\theta_+$ 
and $\theta_-$ can be smaller or larger than $\pi/2$, which corresponds to 
positive or negative contributions to the meniscus height $h$.
Electrowetting can be detected as the dependence of the contact angles 
$\theta_+(U)$ and $\theta_-(U)$, and in turn, via Eq.~\Eq{capillaryrise}, of the 
meniscus height $h(U)$, on the electrostatic potential difference 
$U=\Psi_+-\Psi_-$ applied between the electrodes.


\subsection{Contact angle and effective interface potential}

\begin{figure}[!t]
   \includegraphics{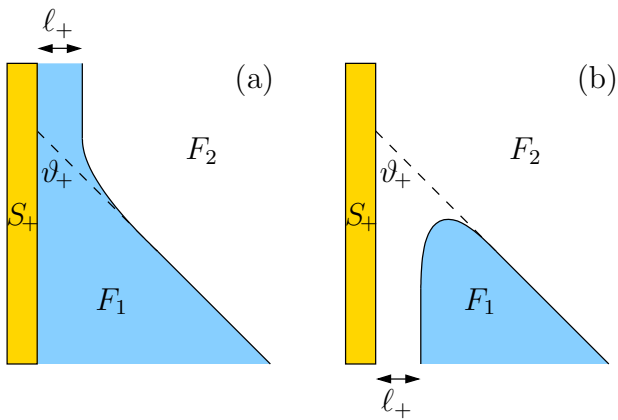}
   \caption{Closeup of the possible geometries of the three-phase contact 
            region formed by the anodic substrate $S_+$ (and similarly for the
            the cathodic substrate $S_-$) and two immiscible fluids
            $F_1$ and $F_2$ marked by the dashed box in Fig.~\ref{fig:1}.
            The fluid which is preferred by the substrates $S_\pm$ is denoted 
            by $A$, whereas the other, less preferred fluid is called $B$.
            Panel (a) corresponds to the case of an $F_1$-philic ($A=F_1$) 
            substrate $S_+$, while panel (b) displays the case of an 
            $F_2$-philic ($A=F_2$) substrate $S_+$.
            Here it is assumed that both substrates $S_+$ and $S_-$ prefer the
            same fluid.
            The macroscopic contact angle $\theta_+$ ($\theta_-$) shown in 
            Fig.~\ref{fig:1} describes the asymptotic inclination of the
            $F_1$-$F_2$ interface far away from the substrate $S_+$ ($S_-$), 
            whereas close to the substrate $S_+$ ($S_-$) a film of microscopic 
            thickness $\ell_+$ ($\ell_-$) of the preferred fluid $A$ is formed.}
   \label{fig:2}
\end{figure}

The contact angles $\theta_+$ and $\theta_-$ in Fig.~\ref{fig:1} provide a 
macroscopic description of the fluid-fluid-substrate three-phase contact region 
(highlighted by the dashed box in Fig.~\ref{fig:1} for the anodic substrate 
$S_+$).
According to the chemical properties of the fluids and the substrates, the 
contact of substrate $S_\pm$ with one fluid, henceforth denoted by fluid
$A$, is more preferable than with the other fluid, henceforth denoted by fluid
$B$.
Here it is assumed that both substrates are chemically equal such that either
fluid $F_1$ or fluid $F_2$ is preferred by both substrates $S_+$ and $S_-$.
Consequently, if substrate $S_\pm$ is macroscopically in contact with the bulk
of the less preferred fluid $B$ and if the thermodynamic state is away from
wetting transitions such that the substrate is only partially wet by phase $A$, 
a film of microscopic extension $\ell_\pm>\xi$ composed of the preferred
fluid $A$ forms in between substrate $S_\pm$ and the bulk of fluid $B$ 
\cite{Dietrich1988}, where $\xi$ denotes the bulk correlation length, which is
of the order of the particle size if the thermodynamic state is away from 
critical points.
Then the fluid structure is similar to that of a composition of an $S_\pm$-$A$ 
interface at the substrate surface and a free $A$-$B$ interface at a distance
$\ell_\pm$ away from the substrate, both being of typical extension $\xi$ 
\cite{Dietrich1988}.
This structure leads to a surface contribution 
$\Omega_{s,\pm B}(\ell_\pm)=\gamma_{\pm A}+\gamma_{12}+\omega_\pm(\ell_\pm)$ to
the grand potential of the system, where $\gamma_{\pm A}$ and $\gamma_{12}$ denote
the $S_\pm$-$A$ and $F_1$-$F_2$ interfacial tensions, respectively, and where
$\omega_\pm(\ell_\pm)$ is the effective interface potential \cite{Dietrich1988}.
It is important to distinguish $\Omega_{s,\pm B}$ from the interfacial tension
$\gamma_{\pm B}$ of an $S_\pm$-$B$ interface in the \emph{absence} of phase $A$.
Here $\Omega_{s,\pm B}\not=\gamma_{\pm B}$ because the presence of the preferred
phase $A$ leads to a structural change, i.e., the formation of $A$-films, as
compared to the situation in the absence of phase $A$.
Ignoring the difference between $\Omega_{s,\pm B}$ and $\gamma_{\pm B}$ is
equivalent to ignoring the formation of $A$-films and it is this crucial 
misconception which underlies the electrocapillarity approach to electrowetting.
In contrast, if substrate $S_\pm$ is in contact with the bulk of the preferred
fluid $A$, the fluid is non-uniform only close to the substrate surfaces up 
to distances $\xi$, and this interfacial structure is not modified
by the presence of fluid $B$, hence $\Omega_{s,\pm A}=\gamma_{\pm A}$.
Depending on whether the preferred fluid $A$ is fluid $F_1$ or fluid $F_2$ the
substrates $S_\pm$ are referred to as $F_1$-philic or $F_2$-philic, respectively.
A closeup of the fluid-fluid-substrate three-phase contact region close to
substrate $S_+$ marked by the dashed box in Fig.~\ref{fig:1} is sketched 
respectively in Figs.~\ref{fig:2}(a) and (b) for an $F_1$-philic ($A=F_1$) 
and an $F_2$-philic ($A=F_2$) substrate.

The macroscopic contact angle $\theta_\pm$ (see Fig.~\ref{fig:2}) is
related to the surface contributions $\Omega{s,\pm 1}$, $\Omega_{s,\pm 2}$
and the interfacial tension $\gamma_{12}$ of the $S_\pm$-$F_1$, $S_\pm$-$F_2$, and 
$F_1$-$F_2$ interface, respectively, by Young's equation 
\cite{Landau2005, Rowlinson2002, DeGennes2004}
\begin{align}
   \Omega_{s,\pm2} = \Omega_{s,\pm1} + \gamma_{12}\cos\theta_\pm.
   \label{eq:Y}
\end{align}
It is common to assume $\Omega_{s,\pm\alpha}=\gamma_{s,\pm\alpha}, 
\alpha\in\{F_1,F_2\}$, but this misconception to ignore the structural
differences of a macroscopic $S_\pm$-$\alpha$ contact in the presence and 
in the absence of additional phases can have significant consequences.
The surface contributions $\Omega_{s,\pm 1}$ and $\Omega_{s,\pm 2}$
are related to the depth of the effective interface potential $\omega_+(\ell)$ 
evaluated at the equilibrium film thickness $\ell=\ell_\pm$ by \cite{Dietrich1988}
\begin{align}
   \Omega_{s,\pm 1} & = \gamma_{\pm 1}
   \notag\\
   \Omega_{s,\pm 2} & = \gamma_{\pm 1} + \gamma_{12} + \omega_\pm(\ell_\pm)
   \label{eq:minimumeipp}
\end{align}
for $F_1$-philic substrates $S_\pm$ (see above the three-phase contact region in
Fig.~\ref{fig:2}(a)) and by
\begin{align}
   \Omega_{s,\pm 1} & = \Omega_{s,\pm 2} + \gamma_{12} + \omega_\pm(\ell_\pm)
   \notag\\
   \Omega_{s,\pm 2} & = \gamma_{s,\pm 2}
   \label{eq:minimumeipm}
\end{align}
for $F_2$-philic substrates $S_\pm$ (see below the three-phase contact region in
Fig.~\ref{fig:2}(b)).
Hence, one obtains \cite{Dietrich1988}
\begin{align}
   \cos\theta_\pm = 
   \frac{\Omega_{s,\pm 2}-\Omega_{s,\pm1}}{\gamma_{12}} =
   p\left(1 + \frac{\omega_\pm(\ell_\pm)}{\gamma_{12}}\right),
   \label{eq:contactangleeipp}
\end{align}
where $p=+1$ for $F_1$-philic and $p=-1$ for $F_2$-philic substrates $S_\pm$.
This equation connects the macroscopic contact angle $\theta_\pm$ with the 
microscopic structure represented by the effective interface potential 
$\omega_\pm(\ell)$ of $A$-films at substrate $S_\pm$ in macroscopic
contact with bulk fluid $B$.

The next Sec.~\ref{subsec:dft} describes an approximate calculation of the
effective interface potentials $\omega_\pm(\ell)$ for the setting of
Fig.~\ref{fig:1}.
The dependence of $\omega_\pm(\ell_\pm;U)$ on the electrostatic potential 
difference $U$ between the electrodes, together with 
Eq.~\Eq{contactangleeipp}, then leads to the 
electrowetting equations derived in Sec.~\ref{subsec:elweteq}.

However, already without explicit expressions for the effective interface potentials,
one can draw an important conceptual conclusion from Eq.~\Eq{contactangleeipp}:
Electrowetting is \emph{not} an electrocapillarity
effect, since \emph{no} $U$-dependent substrate-fluid interfacial tensions,
which describe the contact of the substrate with \emph{one} fluid, occur on the 
right-hand side.
Instead, electrowetting is related to the depth of the effective interface
potential, which describes the $U$-dependence of the microscopic fluid 
structure close to the substrate in the presence of \emph{two} fluids.


\subsection{\label{subsec:dft}Density functional theory of wetting}

\begin{figure}[!t]
   \includegraphics{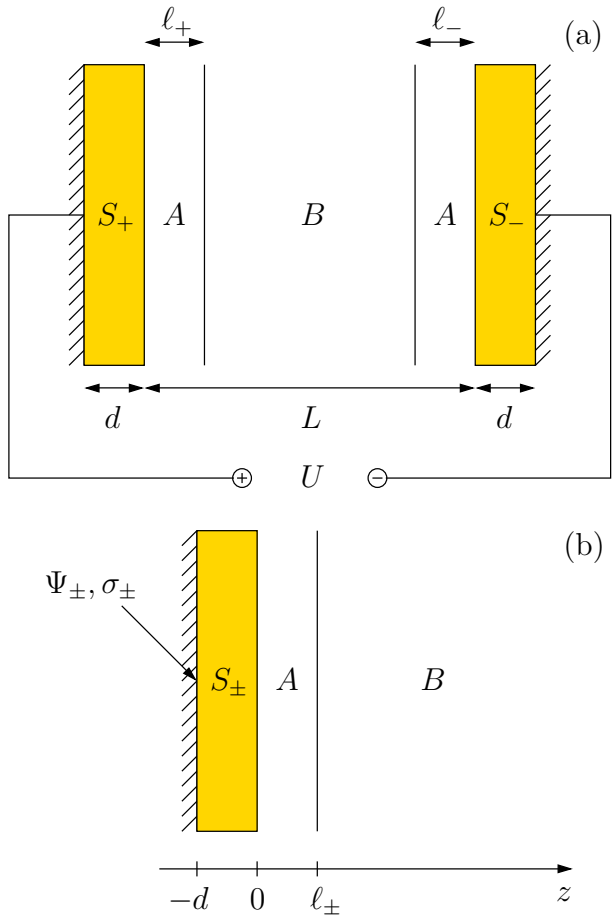}
   \caption{(a) Far above (Fig.~\ref{fig:2}(a)) or below (Fig.~\ref{fig:2}(b)) the
            three-phase contact regions (see Fig.~\ref{fig:1}) the dielectric
            substrates $S_+$ and $S_-$ of thickness $d$ and distance $L$ are 
            covered by films of the preferred fluid $A$ of microscopic 
            thicknesses $\ell_\pm$, which separate the substrates from
            the bulk of the less preferred fluid $B$.
            (b) Since the separation $L$ between the substrates $S_+$ and $S_-$
            is typically the largest length scale, one can consider the limit
            $L\to\infty$, which renders the effective interface potential
            $\omega_\pm(\ell)$ at substrate $S_\pm$ that of a semi-infinite
            system.
            A coordinate axis normal to each substrate is introduced with the
            origin $z=0$ at the surface and the fluids in the range $z>0$. 
            The interface between the $A$-film and the bulk fluid $B$ is located
            at $z=\ell_\pm$ and the electrode is at position $z=-d$, where the
            electrostatic potential is $\Psi_\pm$ and the surface charge density
            is $\sigma_\pm$.}
   \label{fig:3}
\end{figure}

In order to obtain the effective interface potential $\omega_\pm(\ell)$ of
an $A$-film of thickness $\ell$ at substrate $S_\pm$ in Fig.~\ref{fig:1}, 
whose value for the equilibrium film thicknesses $\ell=\ell_\pm$ is related to 
the contact angle $\theta_\pm$ via Eq.~\Eq{contactangleeipp}, one may represent
the structure in Fig.~\ref{fig:1} far 
above (for an $F_1$-philic substrate $S_\pm$, see Fig.~\ref{fig:2}(a)) or 
below (for an $F_2$-philic substrate $S_\pm$, see Fig.~\ref{fig:2}(b)) the 
three-phase contact region by the quasi-one-dimensional slab depicted in 
Fig.~\ref{fig:3}(a).
The chemically identical substrates $S_+$ and $S_-$, which comprise metal 
electrodes coated with dielectric layers of thickness $d$, are separated by a
distance $L$ and covered with films of thicknesses $\ell_+$ and $\ell_-$, 
respectively, of the preferred fluid $A$.
Moreover, the electrodes are assumed to be ideally polarized, i.e.,
electrochemical reactions do not occur.
Even under these conditions the film thicknesses $\ell_+$ and $\ell_-$ can differ,
if unequal partitioning of ions at the film-bulk fluid interfaces takes place.
This is expected to occur in general due to generic differences in solubility
contrasts \cite{Bier2012}.
The macroscopic distance $L$ between the substrates is typically the largest 
length scale such that only the limit $L\to\infty$ is considered in the following.
Hence the effective interface potentials $\omega_+(\ell)$ and $\omega_-(\ell)$ 
at the substrates $S_+$ and $S_-$, respectively, are those of semi-infinite
systems.
For each substrate $S_\pm$ a coordinate axis in normal direction with the origin
$z=0$ at the substrate surface and the fluid at $z>0$ is introduced (see 
Fig.~\ref{fig:3}(b)).
The interface between the $A$-film and the bulk of phase $B$ is located at position
$z=\ell_\pm$ and the electrode is at $z=-d$, where the electrostatic potential
is $\Psi_\pm$ and the surface charge density is $\sigma_\pm$.
 
Considering the two bulk phases of fluids $F_1$ and $F_2$ outside the capacitor in
Fig.~\ref{fig:1} as particle reservoirs, one is naturally led to a grand-canonical 
description of the thermodynamic state.
A starting point for the derivation of the effective interface potential 
$\omega_\pm(\ell)$ is the grand potential functional per thermal energy 
$k_BT=1/\beta$ and per area $\mathcal{A}$ of the electrode
\begin{align}
   \frac{\beta\Omega_1[\phi,\rho_\pm]}{\mathcal{A}} & = 
    \frac{\beta\Omega_0[\phi]}{\mathcal{A}} + 
     \frac{\beta dD(0)^2}{2\epsilon_\text{vac}\epsilon_S} -
     \beta \Psi_\pm D(0) 
   \notag\\
   & +\int\limits_0^\infty\!\!\mathrm{d} z\bigg[\sum_{i=\pm}\rho_i(z)\!
     \bigg(\!\ln\frac{\rho_i(z)}{\zeta_i}-1+\beta V_i(\phi(z))\!\bigg) 
   \notag\\
   & \phantom{+\int\limits_0^\infty\!\!\mathrm{d} z\bigg[}+ 
     \frac{\beta D(z)^2}{2\epsilon_\text{vac}\epsilon(\phi(z))}\bigg] 
   \label{eq:Omega1}
\end{align}
in terms of the solvent composition profile $\phi$ and the $\pm$-ion number 
density profiles $\rho_\pm$.
Here the permittivity $\epsilon_\text{vac}$ of the vacuum, the relative permittivity
$\epsilon_S$ of the substrate $S_\pm$, and the fugacities $\zeta_\pm$ of $\pm$-ions
are used.
The density functional $\Omega_0[\phi]$ describes the grand potential of the pure,
i.e., salt-free fluids.
The electric displacement $D(z)$ is determined by Gau{\ss}'s law and the boundary
condition of global charge neutrality:
\begin{align}
   D'(z)=\sum_{i=\pm}q_ie\rho_i(z), \qquad D(\infty)=0.
   \label{eq:Gauss}
\end{align}
Here $q_\pm=\pm 1$ denotes the valency of $\pm$-ions and $e$ is the elementary charge.
Since the substrate $S_\pm$ is free of ions the electrode surface charge density is 
given by $\sigma_\pm=D(-d^+)=D(0^-)$ (see Fig.~\ref{fig:3}(b)).
In the absence of specific adsorption of ions at the substrate surface, the electric
displacement is continuous at $z=0$, i.e., $D(0^-)=D(0^+)=D(0)$, so that 
$\sigma_\pm=D(0)$.
The electric displacement $D(z)$ is related to the electrostatic potential $\psi(z)$
by $D(z)=-\epsilon_\text{vac}\epsilon(\phi(z))\psi'(z)$.
The electrostatic potential of the electrode is given by $\Psi_\pm=\psi(-d)$ (see 
Fig.~\ref{fig:3}(b)).
The second term on the right-hand side and the term in the last line of 
Eq.~\Eq{Omega1} account for the electrostatic field energy inside the substrate and
the fluids, respectively, whereas the last term in the first line represents the internal
energy of the voltage source sustaining the electrostatic potential $\psi(-d)=\Psi_\pm$
of the electrode (see Fig.~\ref{fig:3}(b)).
The expressions $V_\pm(\phi(z))$ and $\epsilon(\phi(z))$ denote respectively the local
solvation free energy of a $\pm$-ion and the relative permittivity at a position 
$z>0$ where the solvent composition is given by $\phi(z)$ \cite{Bier2012}.
Finally, the second line of Eq.~\Eq{Omega1} describes the grand potential of the ions,
where the ion number densities are assumed to be sufficiently small such that ions 
interact with each other only via the electrostatic field.  

The equilibrium bulk state $(\bar\phi,I)$ with the bulk solvent composition $\bar\phi$
and the bulk ionic strength $I$ minimizes the bulk grand potential density 
$\Omega_b(\bar\phi,I)/\mathcal{V}$, which can be derived from Eq.~\Eq{Omega1} by 
inserting uniform profiles $\phi(z)=\bar\phi$ and $\rho_\pm(z)=I$, omitting all surface
terms, and noting $D=0$ in the bulk.
The two immiscible fluids $A$ and $B$ at coexistence in Fig.~\ref{fig:3}(a) correspond to
two equilibrium bulk states $(\bar\phi_A,I_A)$ and $(\bar\phi_B,I_B)$, respectively,
with equal bulk grand potential density: $\Omega_b(\bar\phi_A,I_A)/\mathcal{V}=
\Omega_b(\bar\phi_B,I_B)/\mathcal{V}$.

Since the present analysis is concerned with films of thicknesses $\ell_\pm>\xi$ but
not with interfacial structures on length scales less than $\xi$ it is natural to
approximate the solvent composition profile $\phi$ in Eq.~\Eq{Omega1} within the 
so-called sharp-kink approximation \cite{Dietrich1988}
\begin{align}
   \phi_\ell(z) :=
   \begin{cases}
      \bar\phi_A, & z<\ell \\
      \bar\phi_B, & z>\ell.
   \end{cases}
   \label{eq:sk}
\end{align}
Furthermore, in view of the small ionic strengths $I_A$ and $I_B$ to be considered
here and for sufficiently small surface potential $|\psi(0)-\psi(\infty)|$, the 
term in the second line of Eq.~\Eq{Omega1} can be expanded up to quadratic order
in the ion number density deviations $\Delta\rho_i(z):=\rho_i(z)-I_\alpha, i=\pm,$ 
with $\alpha=A$ for $z<\ell$ and $\alpha=B$ for $z>\ell$ if $I_\alpha>0$.
This leads to an approximate grand potential functional
\begin{align}
   & \frac{\beta\Omega(\ell,[\Delta\rho_\pm])}{\mathcal{A}} =
   \frac{\beta\Omega_0[\phi_\ell]}{\mathcal{A}} + 
     \frac{\beta dD(0)^2}{2\epsilon_\text{vac}\epsilon_S} -
     \beta \Psi_\pm D(0)
   \notag\\
   & \phantom{MMM}+\int\limits_0^\ell\!\!\mathrm{d} z
     \bigg[f_A(\Delta\rho_+(z),\Delta\rho_-(z)) +
           \frac{\beta D(z)^2}{2\epsilon_\text{vac}\epsilon_A}\bigg] 
   \notag\\
   & \phantom{MMM}+\int\limits_\ell^\infty\!\!\mathrm{d} z
     \bigg[f_B(\Delta\rho_+(z),\Delta\rho_-(z)) +
           \frac{\beta D(z)^2}{2\epsilon_\text{vac}\epsilon_B}\bigg]
   \label{eq:Omega}
\end{align}
with 
\begin{align}
   & f_\alpha(\Delta\rho_+,\Delta\rho_+) = \label{eq:falpha}\\
   & \begin{cases}
   \dps\sum_{i=\pm}\!\bigg[I_\alpha\Big(\ln\frac{I_\alpha}{\zeta_i}-1+\beta V_i(\bar\phi_\alpha)\Big)  & \\
   \dps\phantom{\sum_{i=\pm}\!\bigg[}+ \Big(\ln\frac{I_\alpha}{\zeta_i}+\beta V_i(\bar\phi_\alpha)\Big)\Delta\rho_i
     +\frac{\Delta\rho_i^2}{2I_\alpha}\!\bigg], & I_\alpha>0 \\
   0, & I_\alpha = 0.
   \end{cases}
   \notag
\end{align}

By minimizing $\beta\Omega(\ell,[\Delta\rho_\pm])/\mathcal{A}$ in 
Eq.~\Eq{Omega} with respect to the profiles $\Delta\rho_\pm$ one
obtains the equilibrium profiles $\Delta\rho_\pm^{(\ell)}$.
Inserting $\Delta\rho_\pm^{(\ell)}$ into Eq.~\Eq{Omega} and subtracting
the bulk contribution $\beta\Omega_b(\bar\phi_B,I_B)/\mathcal{A}$ leads to the
surface contribution to the grand potential \cite{Dietrich1988}
\begin{align}
   \Omega_s(\ell) =
   \frac{\Omega(\ell,[\Delta\rho_\pm^{(\ell)}]) -
   \Omega_b(\bar\phi_B,I_B)}{\mathcal{A}}.
   \label{eq:Omegas}
\end{align}
Finally, the effective interface potential at substrate $S_\pm$ is given by 
$\omega_\pm(\ell)=\Omega_s(\ell)-\Omega_s(\infty)$ \cite{Dietrich1988}, 
which, in the present context, can be written in the form
\begin{align}
   \omega_\pm(\ell) = &
   \omega_0(\ell) 
   + \frac{A(\ell)}{2}(\Psi_\pm-\bar\psi_A)^2 
   \notag\\
   & + B(\ell)(\Psi_\pm-\bar\psi_A) + C(\ell).
   \label{eq:omega}
\end{align}
Here $\omega_0(\ell)$ denotes the effective interface potential corresponding
to the grand potential functional $\Omega_0[\phi]$ of the pure fluids, 
$\bar\psi_\alpha:=-(k_BT\ln(I_\alpha/\zeta_+)+V_+(\bar\phi_\alpha))/e=
(k_BT\ln(I_\alpha/\zeta_-)+V_-(\bar\phi_\alpha))/e$ is the Galvani potential 
of phase $\alpha\in\{A,B\}$, and
\begin{align}
   A(\ell) &= \frac{\epsilon_\text{vac}}{Q(\ell)\lambda_B}\
              \frac{\lambda_B-\lambda_A}{\lambda_S+\lambda_A}
              \exp\Big(-\frac{\lambda_\ell}{\lambda_A}\Big) 
              \label{eq:coefficients}\\
   B(\ell) &= -\frac{\epsilon_\text{vac}}{Q(\ell)\lambda_B}
              (\bar\psi_A-\bar\psi_B) 
              \notag\\
   C(\ell) &= \frac{\epsilon_\text{vac}}{2Q(\ell)\lambda_B}\
              \frac{\lambda_S-\lambda_A}{\lambda_B+\lambda_A}
              \exp\Big(-\frac{\lambda_\ell}{\lambda_A}\Big) 
              (\bar\psi_A-\bar\psi_B)^2 
              \notag\\
   Q(\ell) &= \cosh\Big(\frac{\lambda_\ell}{\lambda_A}\Big)
              \Big(1 + \frac{\lambda_S}{\lambda_B}\Big) +
              \sinh\Big(\frac{\lambda_\ell}{\lambda_A}\Big)
              \Big(\frac{\lambda_A}{\lambda_B} + \frac{\lambda_S}{\lambda_A}\Big)
              \notag
\end{align}
with the length scales $\lambda_A:=1/(\kappa_A\epsilon_A)$, 
$\lambda_B:=1/(\kappa_B\epsilon_B)$, $\lambda_S:=d/\epsilon_S$, and 
$\lambda_\ell:=\ell/\epsilon_A$, where 
$\kappa_\alpha^2:=2\beta e^2I_\alpha/(\epsilon_\text{vac}\epsilon_\alpha)$
is the square of the inverse Debye length in the bulk of phase 
$\alpha\in\{A,B\}$.
In addition, the electrode charge density can be written as
\begin{align}
   \sigma_\pm = D(0) = F(\ell_\pm)(\Psi_\pm-\bar\psi_A) - B(\ell_\pm)
   \label{eq:sigma}
\end{align}
with
\begin{align}
   F(\ell) = 
   \frac{\epsilon_\text{vac}}{Q(\ell)}
   \Big(\frac{1}{\lambda_A}\sinh\Big(\frac{\lambda_\ell}{\lambda_A}\Big) +
   \frac{1}{\lambda_B}\cosh\Big(\frac{\lambda_\ell}{\lambda_A}\Big)\Big)
   \label{eq:F}
\end{align}
The Galvani potential difference (Donnan potential) between the 
phases $A$ and $B$, $\bar\psi_A-\bar\psi_B=((V_-(\bar\phi_A)-V_+(\bar\phi_A))-
(V_-(\bar\phi_B)-V_+(\bar\phi_B)))/(2e)$, can be inferred from 
the solubility of the $\pm$-ions in the solvents $A$ and $B$ 
\cite{Lyklema1991, Inerowicz1994, Kalidas2000, Marcus2007}.
Moreover, $\Psi_\pm-\bar\phi_A$ is determined by the potential difference 
$U=\Psi_+-\Psi_-$ and the fact that no chemical reactions take
place at the electrodes so that the total charge of both electrodes vanishes:
$\sigma_++\sigma_-=0$.
Using Eq.~\Eq{sigma} leads to
\begin{align}
   \Psi_+-\bar\psi_A =& 
   \frac{F(\ell_-)U + B(\ell_+) + B(\ell_-)}{F(\ell_+) + F(\ell_-)} 
   \notag\\
   \Psi_--\bar\psi_A =& 
   \frac{-F(\ell_+)U + B(\ell_+) + B(\ell_-)}{F(\ell_+) + F(\ell_-)} 
   \label{eq:PsimpsiA}
\end{align}


\subsection{\label{subsec:elweteq}Electrowetting equation}

The equilibrium film thicknesses $\ell_+$ and $\ell_-$ are both similar in 
magnitude ($\ell_+\approx\ell_-$) of the order of a few bulk correlation 
lengths $\xi$ away from wetting transitions.
Hence, Eq.~\Eq{PsimpsiA} leads to $\Psi_\pm-\bar\psi_A\approx\pm U/2+
(B(\ell_+)+B(\ell_-))/(F(\ell_+)+F(\ell_-))$, i.e., the $U$-dependent part
of $\Psi_\pm-\bar\psi_A$ is rather insensitive to a variation of 
$\ell_+\approx\ell_-$.
Moreover, the film thicknesses $\ell_+$ and $\ell_-$ are typically much smaller
than the Debye length $1/\kappa_A$ in the $A$-film.
Consequently, the leading $U$-dependent contribution $\sim U^2$ to the 
effective interface potential $\omega_\pm(\ell)$ in Eq.~\Eq{omega}, which
decays exponentially with $\ell$ on the length scale of half of a 
Debye length, $1/(2\kappa_A)$ (see Eq.~\Eq{coefficients}), does not 
significantly shift the equilibrium film thickness $\ell_\pm$ but it merely
lifts the depth $\omega_\pm(\ell_\pm;U)$ of the effective interface potential.
Therefore, in the following, the film thicknesses $\ell_\pm$ are considered as
independent of the applied voltage $U$.

Hence, Eq.~\Eq{contactangleeipp} can be written as
\cite{Mugele2005}
\begin{align}
   \cos\theta_\pm(U) = \cos\theta_\pm(0) + \eta_\pm(U)
   \label{eq:contactangleeip}
\end{align}
with the electrowetting number
\begin{align}
   \eta_\pm(U) := p\frac{\omega_\pm(\ell_\pm;U)-\omega_\pm(\ell_\pm;0)}{\gamma_{12}}.
   \label{eq:ewn1}
\end{align}
Inserting Eq.~\Eq{PsimpsiA} into Eq.~\Eq{omega} leads to 
\begin{align}
   \eta_\pm(U) = 
   & \frac{p}{\gamma_{12}}\bigg(\frac{A(\ell_\pm)F(\ell_\mp)^2}{2(F(\ell_+)+F(\ell_-))^2}U^2 
   \notag\\
   & \phantom{\frac{p}{\gamma_{12}}\Big(}
     \pm\Big(\frac{A(\ell_\pm)F(\ell_\mp)(B(\ell_+)+B(\ell_-))}{(F(\ell_+)+F(\ell_-))^2}
   \notag\\
   & \phantom{\frac{p}{\gamma_{12}}\bigg(\pm\Big(}
     + \frac{B(\ell_\pm)F(\ell_\mp)}{F(\ell_+)+F(\ell_-)}\Big)U\bigg).
   \label{eq:ewn2}
\end{align}
This equation is expected to be valid for sufficiently small voltages $|U|$
such that the quadratic approximation Eq.~\Eq{falpha} is applicable.
The electrowetting number $\eta_\pm(U)$ in Eq.~\Eq{ewn2} differs from 
those in the literature \cite{Mugele2005} in a number of aspects, as is
discussed in the next section.
The most obvious difference is the occurrence of a correction term
$\sim U$, which vanishes exactly only if $B(\ell)=0$ due to a
vanishing Donnan potential (Galvani potential difference) 
$\bar\psi_A-\bar\psi_B$.
For $\bar\psi_A-\bar\psi_B\not=0$, i.e., $B(\ell)\not=0$, the 
electrowetting number $\eta_\pm(U)$ in Eq.~\Eq{ewn2} is not minimal at and
not symmetric with respect to $U=0$.
However, for a sufficiently large voltage $|U|$ the subleading term $\sim U$ 
is dominated by the leading term $\sim U^2$.

Whereas the full expression for the electrowetting number $\eta_\pm(U)$ in 
Eq.~\Eq{ewn2} depends on the five possibly largely different length scales 
$\lambda_A$, $\lambda_B$, $\lambda_S$, $\lambda_{\ell_+}$, and 
$\lambda_{\ell_-}$, the latter two, corresponding to the thicknesses of the 
$A$-films at the substrates $S_+$ and $S_-$, respectively, are typically
of similar magnitude: $\ell_+\approx\ell_-$, i.e., 
$\lambda_{\ell_+}\approx\lambda_{\ell_-}$.
This case $\ell_+=\ell_-=:\ell$ is discussed in detail in the next section, for which
the electrowetting number $\eta_\pm(U)$ in Eq.~\Eq{ewn2} simplifies to
\begin{align}
   \eta_\pm(U) = 
   & \frac{p}{\gamma_{12}}\bigg(\frac{A(\ell)}{8}U^2 
     \pm\frac{B(\ell)}{2}\Big(\frac{A(\ell)}{F(\ell)}+1\Big)U\bigg).
   \label{eq:ewn3}
\end{align}
Moreover, the film thicknesses $\ell_+$ and $\ell_-$ are typically smaller than
the Debye lengths $1/\kappa_A$ and $1/\kappa_B$ so that the  limiting case
$\lambda_\ell\ll\lambda_A,\lambda_B$ is considered throughout, within which
\begin{align}
   A(\ell) & \simeq \frac{\epsilon_\text{vac}}{Q(\ell)}\ 
                    \frac{\lambda_B - \lambda_A}{\lambda_B(\lambda_S + \lambda_A)}
   \label{eq:Aell}\\
   B(\ell) & = -\frac{\epsilon_\text{vac}}{Q(\ell)}\ 
               \frac{\bar\psi_A-\bar\psi_B}{\lambda_B}
   \label{eq:Bell}\\
   F(\ell) & \simeq \frac{\epsilon_\text{vac}}{Q(\ell)}
                    \left(\frac{\lambda_\ell}{\lambda_A^2} + \frac{1}{\lambda_B}\right)
   \label{eq:Fell}\\
   Q(\ell) & \simeq 1 + \frac{\lambda_S}{\lambda_B} + 
                    \frac{\lambda_\ell\lambda_S}{\lambda_A^2}.
   \label{eq:Qell}                         
\end{align}


\section{\label{sec:discussion}Discussion}

\subsection{Electrowetting and electrocapillarity}

Before discussing the electrowetting number in Eq.~\Eq{ewn3}
obtained within the present density functional analysis, the traditional
approach based on the assumption of electrowetting being an electrocapillarity
effect \cite{Pellat1894, Pellat1895, Sparnaay1964, Sondag-Huethorst1992, 
Berge1993, Vallet1996, Welters1998, Blake2000, Decamps2000, Quinn2003, 
Jones2004, Mugele2005, Quinn2005, Klarman2011, Sedev2011, Daub2012} is 
repeated. 
Here only the classical method based on Lippmann's equation is 
presented.
However, calculations using alternative methods, e.g., based on Maxwell's
stress tensor \cite{Kang2002, Mugele2005}, suffer from the same misconceptions.

The starting point is Young's equation~\Eq{Y} but with the 
\emph{incorrect} assumption $\Omega_{s,\pm\alpha}=\gamma_{\pm\alpha}, 
\alpha\in\{F_1,F_2\}$.
In order to obtain the $U$-dependence of the interfacial tension 
$\gamma_{\pm\alpha}$ one considers a semi-infinite fluid $\alpha$ bound by
a planar substrate $S_\pm$.
The interfacial tension $\gamma_{\pm\alpha}$ changes upon changing the
electrostatic potential $\psi_{\pm\alpha}$ of substrate $S_\pm$ with
respect to that of the bulk of phase $\alpha$ according to Lippmann's
equation \cite{Lippmann1875,Grahame1947}
\begin{align}
   \frac{\partial\gamma_{\pm\alpha}}{\partial\psi_{\pm\alpha}}
   = -\sigma_{\pm\alpha},
   \label{eq:Lippmann}
\end{align}
where $\sigma_{\pm\alpha}$ is the surface charge density of substrate $S_\pm$
in contact with phase $\alpha$.
Describing the $S_\pm$-$\alpha$ interface by means of the
potential-independent differential capacitance 
$C_{S\alpha}=\partial\sigma_{\pm\alpha} / \partial\psi_{\pm\alpha}$, 
which is assumed to not depend on $S_\pm$ for chemically identical
substrates, and integrating twice with respect to the electrostatic substrate 
potential $\psi_{\pm\alpha}$ using Lippmann's equation~\Eq{Lippmann} leads to
\begin{align}
   \gamma_{\pm\alpha}(\psi_{\pm\alpha}) = 
   \gamma_{\pm\alpha}(0) - 
   \frac{C_{S\alpha}}{2}\psi_{\pm\alpha}^2.
   \label{eq:intLipp}
\end{align}
Young's equation~\Eq{Y} in conjunction with the assumption 
$\Omega_{s,\pm\alpha}=\gamma_{\pm\alpha}$ reads
\begin{align}
   \gamma_{\pm 2}(\psi_{\pm 2})
   & = \gamma_{\pm 2}(0) - \frac{C_{S2}}{2}\psi_{\pm 2}^2 
   \notag\\
   & = \gamma_{\pm 1}(\psi_{\pm 1}) + \gamma_{12}\cos\theta_\pm
   \notag\\ 
   &= \gamma_{\pm 1}(0) - \frac{C_{S1}}{2}\psi_{\pm 1}^2
      + \gamma_{12}\cos\theta_\pm.
\end{align}
Noting $\gamma_{\pm 2}(0)-\gamma_{\pm 1}(0) = \gamma_{12}\cos\theta_\pm(0)$
leads to 
\begin{align}
   \cos\theta_\pm - \cos\theta_\pm(0) 
   = \frac{C_{S1}}{2\gamma_{12}}\psi_{\pm 1}^2
       - \frac{C_{S2}}{2\gamma_{12}}\psi_{\pm 2}^2.
   \label{eq:eweec1}
\end{align}
Using $\sigma_{\pm\alpha}=C_{S\alpha}\psi_{\pm\alpha}$, one obtains the
analog of Eq.~\Eq{PsimpsiA} from $U=\psi_{+\alpha}-\psi_{-\alpha}$ and 
$\sigma_{+\alpha}+\sigma_{-\alpha}=0$ as $\psi_{\pm\alpha} = \pm U/2$.
This leads to the commonly used form of the electrowetting equation 
\cite{Pellat1894, Pellat1895, Sparnaay1964, Sondag-Huethorst1992, Berge1993,
Vallet1996, Welters1998, Blake2000, Decamps2000, Quinn2003, Jones2004,
Mugele2005, Quinn2005, Klarman2011, Sedev2011, Daub2012}
\begin{align}
   \cos\theta_\pm(U) - \cos\theta_\pm(0) =
   \frac{C_{S1}-C_{S2}}{8\gamma_{12}}U^2 =:
   \eta_\text{ec}(U)
   \label{eq:eweec2}
\end{align}
with the differential capacitances $C_{S\alpha}$ being those of a substrate 
in macroscopic contact with only one fluid phase $\alpha$.
These differential capacitances $C_{S\alpha}$ can typically be interpreted as 
those of a capacitor of capacitance $C_S=\epsilon_\text{vac}\epsilon_S/d$, 
representing substrate $S_\pm$, connected \emph{in series} with a capacitor 
of capacitance $C_\alpha$, representing fluid $\alpha$: 
$1/C_{S\alpha}=1/C_S+1/C_\alpha$.
If fluid $\alpha$ is an electrolyte solution the fluid capacitance is that of the
electric double layer in a semi-infinite system, 
$C_\alpha=\epsilon_\text{vac}\kappa_\alpha\epsilon_\alpha$, whereas for a 
non-conducting dielectric fluid 
$C_\alpha=\lim_{L\to\infty}\epsilon_\text{vac}\epsilon_\alpha/L=0$.
Using the length scales defined after Eq.~\Eq{coefficients}, this leads to
\begin{align}
   C_{S\alpha} =
   \begin{cases}
      \dps\frac{\epsilon_\text{vac}}{\lambda_S+\lambda_\alpha}, 
      & \text{$\alpha$ electrolyte solution} \\
      0,
      & \text{$\alpha$ non-conducting fluid.}
   \end{cases}
   \label{eq:CSalpha}
\end{align}

Equations~\Eq{eweec2} and \Eq{CSalpha} represent the interpretation of 
electrowetting as an electrocapillarity effect \cite{Mugele2005}.
However, the crucial misconception underlying this interpretation is to
use the approximation $\Omega_{s,\pm\alpha}=\gamma_{\pm\alpha}$ and hence
the differential capacitance $C_{S\alpha}$, which corresponds to a
semi-infinite system of a single phase $\alpha$ bound by substrate $S_\pm$,
instead of accounting for the actual fluid structure at the substrate.
The interfacial structure, and therefore surface quantities such as the 
surface contribution to the grand potential as well as the differential 
capacitance, of substrate $S_\pm$ in macroscopic contact with the bulk fluid
$B$ depend significantly on whether the preferred fluid $A$ is present or not
because an $A$-film forms in between the substrate $S_\pm$ and the bulk fluid
$B$ in the former case whereas it does not in the latter case.
In contrast, these structural properties are naturally accounted for within the
present density functional approach, which relates the contact angle to the
effective interface potential (see Eq.~\Eq{contactangleeipp}), a quantity
which correctly describes the contact of a substrate with \emph{both} fluids
$A$ and $B$.


\subsection{Electrowetting on uncoated metal electrodes}

The early investigations of electrocapillarity by Lippmann \cite{Lippmann1875}
and Pellat \cite{Pellat1894, Pellat1895} have been performed for metal 
electrodes without any dielectric coating. 
At that time for some decades mercury electrodes became the experimental 
standard for investigations of the electric double layer \cite{Grahame1947}.
Pure metal electrodes can be considered as substrates with thickness $d$ being
the smallest length scale: $\lambda_S\ll\lambda_\ell\ll\lambda_A,\lambda_B$.
Within this limit one obtains $Q(\ell)\simeq1$ from Eq.~\Eq{Qell} and 
subsequently from Eqs.~\Eq{Aell}--\Eq{Fell}
\begin{align}
   A(\ell) & \simeq 
   \epsilon_\text{vac}\left(\frac{1}{\lambda_A} - \frac{1}{\lambda_B}\right)
   \\
   B(\ell) & \simeq
   -\frac{\epsilon_\text{vac}}{\lambda_B}(\bar\psi_A-\bar\psi_B)
   \\
   F(\ell) & \simeq
   \epsilon_\text{vac}\left(\frac{\lambda_\ell}{\lambda_A^2} + 
   \frac{1}{\lambda_B}\right).
\end{align}

For the case $\lambda_A\ll\lambda_B$, which is typically the case for
hydrophilic substrates, an aqueous electrolyte solution $F_1=A$ (i.e.,
$p=+1$), and an oil $F_2=B$, one obtains for the the electrowetting number 
Eq.~\Eq{ewn3}
\begin{align}
   \eta_\pm(U) 
   & \simeq \frac{\epsilon_\text{vac}}{8\gamma_{12}\lambda_A}U^2 
            \mp\frac{\epsilon_\text{vac}(\bar\psi_A-\bar\psi_B)}
                    {2\gamma_{12}(\lambda_A+\lambda_\ell\lambda_B/\lambda_A)}U
   \notag\\
   & \simeq \frac{\epsilon_\text{vac}}{8\gamma_{12}\lambda_A}U^2, \quad
     \text{for $|U|\gg 4|\bar\psi_A-\bar\psi_B|$.}
   \label{eq:ewnmetalphil}
\end{align}
Hence, if the voltage $|U|$ is much larger than the Donnan potential (Galvani
potential difference) $|\bar\psi_A-\bar\psi_B|$, the electrowetting number 
$\eta_\pm(U)$ agrees with that in Eq.~\Eq{eweec2}, where 
$C_{S1}\simeq\epsilon_\text{vac}/\lambda_A, C_{S2}=0$ due to Eq.~\Eq{CSalpha}.

Similarly, for the case $\lambda_A\gg\lambda_B$, which is typically the case
for hydrophobic substrates, an oil $F_2=A$ (i.e., $p=-1$), and an 
aqueous electrolyte solution $F_1=B$, one obtains for the electrowetting 
number Eq.~\Eq{ewn3}
\begin{align}
   \eta_\pm(U) 
   & \simeq \frac{\epsilon_\text{vac}}{8\gamma_{12}\lambda_B}U^2 
            \pm\frac{\epsilon_\text{vac}(\bar\psi_A-\bar\psi_B)}
                    {2\gamma_{12}\lambda_A}U
   \notag\\
   & \simeq \frac{\epsilon_\text{vac}}{8\gamma_{12}\lambda_B}U^2, \quad
     \text{for $|U|\gg 4|\bar\psi_A-\bar\psi_B|$.}
   \label{eq:ewnmetalphob}
\end{align}
Again, if the voltage $|U|$ is much larger than the Donnan potential (Galvani
potential difference) $|\bar\psi_A-\bar\psi_B|$, the electrowetting number
$\eta_\pm(U)$ again agrees with that in Eq.~\Eq{eweec2}, where 
$C_{S1}\simeq\epsilon_\text{vac}/\lambda_B, C_{S2}=0$ due to Eq.~\Eq{CSalpha}.

Therefore, the present formalism (Eqs.~\Eq{ewn3}--\Eq{Qell}) confirms the
electrocapillarity-based form of the electrowetting number for the case of 
uncoated metal electrodes ($\eta_\pm(U) \simeq \eta_\text{ec}(U)$), provided
the voltage $|U|$ is sufficiently large as compared to the Donnan potential
(Galvani potential difference) $|\bar\psi_A-\bar\psi_B|$.
Interestingly, for uncoated metal electrodes it is irrelevant whether they
are $F_1$-philic (hydrophilic) or $F_2$-philic (hydrophobic).

However, a small voltage $|U|\ll|\bar\psi_A-\bar\psi_B|$ or 
$\lambda_A\approx\lambda_B$, e.g., for two immiscible electrolyte solutions, 
leads to electrowetting numbers $\eta_\pm(U)\sim U$, in contrast to 
$\eta_\text{ec}(U)\approx0$ in Eq.~\Eq{eweec2} due to $C_{S1}
\approx C_{S2}$ according to Eq.~\Eq{CSalpha}.
Since these conditions are rather special, this scenario is not expected to 
be of practical relevance, but it might provide a test for the present approach.


\subsection{\label{sec:phob}Electrowetting of water on hydrophobic dielectrics 
in oil}

In the last few decades most electrowetting settings used electrodes coated
with an isolating dielectric for technical advantage \cite{Berge1993}.
Almost all of these studies used drops of an aqueous electrolyte solution 
$F_1$ placed on a hydrophobic dielectric and an oil $F_2$ as the environmental
fluid in order to achieve large contact angle ranges being covered by 
electrowetting \cite{Mugele2005}.
Within the present notation this situation corresponds to $A=F_2$ (i.e.,
$p=-1$) and $B=F_1$.
Since the thickness $\ell$ of the microscopic oil film on the substrates 
$S_\pm$ is typically smaller than the the Debye length $1/\kappa_B$ of the 
dilute electrolyte solution $B=F_1$, which in turn is typically much smaller
than the thickness $d$ of the dielectric substrates $S_\pm$, one 
identifies the case $\lambda_\ell\ll\lambda_B\ll\lambda_S\ll\lambda_A$,
where a (practically) ion-free oil $A=F_2$ ($I_A\approx0$) is assumed.
For this regime Eqs.~\Eq{Aell}--\Eq{Qell} read
\begin{align}
   Q(\ell) & \simeq \frac{\lambda_S}{\lambda_B}
   \\
   A(\ell) & \simeq 
   -\frac{\epsilon_\text{vac}}{\lambda_S}
   \\
   B(\ell) & \simeq
   -\frac{\epsilon_\text{vac}}{\lambda_S}(\bar\psi_A-\bar\psi_B)
   \\
   F(\ell) & \simeq
   \frac{\epsilon_\text{vac}}{\lambda_S}
\end{align}
and hence Eq.~\Eq{ewn3} is given by
\begin{align}
   \eta_\pm(U) 
   & \simeq \frac{\epsilon_\text{vac}}{8\gamma_{12}\lambda_S}U^2 
            \pm\frac{\epsilon_\text{vac}(\bar\psi_A-\bar\psi_B)}
                    {2\gamma_{12}\lambda_A}U
   \label{eq:phobwall}\\
   & \simeq \frac{\epsilon_\text{vac}}{8\gamma_{12}\lambda_S}U^2, \quad
     \text{for $|U|\gg 4\frac{\lambda_S}{\lambda_A}|\bar\psi_A-\bar\psi_B|$.}
   \notag
\end{align}
Since $\lambda_S/\lambda_A\ll1$, the approximation in the second line of
the previous equation almost always applies.
It shows that the electrowetting number $\eta_\pm(U)$ for water on a 
hydrophobic dielectric in oil is also in agreement with 
$\eta_\text{ec}(U)$ in Eq.~\Eq{eweec2} with 
$C_{S1}\simeq\epsilon_\text{vac}/\lambda_S, C_{S2}=0$ due to Eq.~\Eq{CSalpha}.


\subsection{\label{sec:phil}Electrowetting of water on hydrophilic dielectrics 
in oil}

Replacing the hydrophobic dielectric substrate of the previous 
Sec.~\ref{sec:phob} by a hydrophilic one leads to the case
$A=F_1$ (i.e., $p=+1$), $B=F_2$, and 
$\lambda_\ell\ll\lambda_A\ll\lambda_S\ll\lambda_B$, where again a 
(practically) ion-free oil $B=F_2$ ($I_B\approx0$) is assumed.
For this regime Eqs.~\Eq{Aell}--\Eq{Qell} read
\begin{align}
   Q(\ell) & \simeq 1 + \frac{\lambda_\ell\lambda_S}{\lambda_A^2}
   = 1 + \frac{\epsilon_A}{\epsilon_S}\ \kappa_A\ell\ \kappa_A d
   \label{eq:Qcorr}\\
   A(\ell) & \simeq \frac{\epsilon_\text{vac}}{Q(\ell)}\ \frac{1}{\lambda_S}
   \\
   B(\ell) & = -\frac{\epsilon_\text{vac}}{Q(\ell)}\ 
               \frac{\bar\psi_A-\bar\psi_B}{\lambda_B}
   \\
   F(\ell) & \simeq
   \frac{\epsilon_\text{vac}}{Q(\ell)}
   \left(\frac{\lambda_\ell}{\lambda_A^2}+\frac{1}{\lambda_B}\right),
\end{align}
which leads to Eq.~\Eq{ewn3} of the form
\begin{align}
   \eta_\pm(U) 
   & \simeq \frac{1}{Q(\ell)}\bigg(
            \frac{\epsilon_\text{vac}}{8\gamma_{12}\lambda_S}U^2 
            \mp\frac{\epsilon_\text{vac}(\bar\psi_A-\bar\psi_B)}
                    {2\gamma_{12}\lambda_B}
   \label{eq:philwall}\\
   & \phantom{\simeq \frac{1}{Q(\ell)}\bigg(}
     \times\Big(\frac{1}{\lambda_S(\lambda_\ell/\lambda_A^2+1/\lambda_B)}+1\Big)U\bigg)
   \notag\\
   & \simeq \frac{1}{Q(\ell)}\ \frac{\epsilon_\text{vac}}{8\gamma_{12}\lambda_S}U^2, \quad
     \text{for $|U|\gg 4|\bar\psi_A-\bar\psi_B|$.}
   \notag
\end{align}

Within the electrocapillarity approach one again expects, as in the previous 
Sec.~\ref{sec:phob}, an electrowetting number $\eta_\text{ec}(U) =
\epsilon_\text{vac}U^2/(8\gamma_{12}\lambda_S)$ (see Eqs.~\Eq{eweec2}
and \Eq{CSalpha}).
However, the electrowetting number $\eta_\pm(U)$ within the present density
functional approach in Eq.~\Eq{philwall}, for sufficiently large voltage 
$|U|\gg4|\bar\psi_A-\bar\psi_B|$, is actually smaller than $\eta_\text{ec}(U)$
by a factor $1/Q(\ell)$: $\eta_\pm(U)\simeq\eta_\text{ec}(U)/Q(\ell)$.

It is apparent from Eq.~\Eq{Qcorr} that $Q(\ell)$ is 
\emph{not} necessarily close to unity, because the typically small value
$\kappa_A\ell\ll 1$ is multiplied with the typically large value
$\kappa_A d\epsilon_A/\epsilon_S\gg 1$.
Assuming typical values of, e.g., dielectric layers of thicknesses 
$d=1\;\mathrm{\mu m}$ and dielectric constant $\epsilon_S=2$, a Debye 
length $1/\kappa_A=10\;\mathrm{nm}$ in the aqueous ($\epsilon_A=80$) 
electrolyte solution $F_1=A$, and thicknesses $\ell=1\;\mathrm{nm}$ of the 
electrolyte films on the substrates, Eq.~\Eq{Qcorr} leads to 
$Q(\ell)\approx400$.
Hence, for this example of electrowetting on a hydrophilic dielectric, the 
analysis leads to electrowetting numbers $\eta_\pm(U)$ which are more than 
two orders of magnitude smaller than expected within the electrocapillarity
approach: $\eta_\pm(U)\approx 0.0025 \eta_\text{ec}(U) \ll \eta_\text{ec}(U)$.

It appears as if no experimental studies of electrowetting on hydrophilic 
substrates have been reported so far.
This is remarkable since the preparation of hydrophilic substrates is
standard in surface science.


\subsection{Electrowetting of immiscible electrolyte solutions on dielectrics}

Whereas the previous two Secs.~\ref{sec:phob} and \ref{sec:phil} considered
an electrolyte solution and an oil, here the case of two immiscible 
electrolyte solutions is discussed.
This situation is characterized by $\lambda_\ell \ll \lambda_A,\lambda_B \ll
\lambda_S$, which leads to Eqs.~\Eq{Aell}--\Eq{Qell} of the form
\begin{align}
   Q(\ell) & \simeq \frac{\lambda_S}{\lambda_A}
   \left(\frac{\lambda_A}{\lambda_B} + \frac{\lambda_\ell}{\lambda_A}\right)
   = \frac{\epsilon_B}{\epsilon_S}\ \kappa_B d + \frac{\epsilon_A}{\epsilon_S}\ \kappa_A\ell\ \kappa_A d
   \label{eq:Qcorri}\\
   A(\ell) & \simeq \frac{\epsilon_\text{vac}}{Q(\ell)}\ \frac{\lambda_B-\lambda_A}{\lambda_S\lambda_B}
   \\
   B(\ell) & = -\frac{\epsilon_\text{vac}}{Q(\ell)}\ 
               \frac{\bar\psi_A-\bar\psi_B}{\lambda_B}
   \\
   F(\ell) & \simeq
   \frac{\epsilon_\text{vac}}{Q(\ell)}
   \left(\frac{\lambda_\ell}{\lambda_A^2}+\frac{1}{\lambda_B}\right).
\end{align}

If electrolyte solutions $F_1$ and $F_2$ are defined by 
$\lambda_{F_1}\leq\lambda_{F_2}$, i.e., $\epsilon_{F_1}I_{F_1}\geq
\epsilon_{F_2}I_{F_2}$, the following three cases have to be distinguished:
(i) $A=F_1$ (i.e., $p=+1$) and $B=F_2$ with $\lambda_A\ll\lambda_B$,
(ii) $A=F_2$ (i.e., $p=-1$) and $B=F_1$ with $\lambda_A\gg\lambda_B$, and
(iii) $\lambda_A\approx\lambda_B$.

Case (i) leads to the electrowetting number Eq.~\Eq{ewn3}
\begin{align}
   \eta_\pm(U) 
   & \simeq \frac{1}{Q(\ell)}\bigg(
            \frac{\epsilon_\text{vac}}{8\gamma_{12}\lambda_S}U^2 
            \mp\frac{\epsilon_\text{vac}(\bar\psi_A-\bar\psi_B)}
                    {2\gamma_{12}\lambda_B}U\bigg)
   \label{eq:ewnimmisciblei}\\
   & \simeq \frac{1}{Q(\ell)}\ \frac{\epsilon_\text{vac}}{8\gamma_{12}\lambda_S}U^2, \quad
     \text{for $|U|\gg 4\frac{\lambda_S}{\lambda_B}|\bar\psi_A-\bar\psi_B|$.}
   \notag
\end{align}
Hence $\eta_\pm(U)\simeq\eta_\text{ec}/Q(\ell)$, where, however, the
depression factor $1/Q(\ell)$ here is typically much smaller than that
of the previous Sec.~\ref{sec:phil} because typically $\epsilon_B\kappa_B 
d/\epsilon_S \gg 1$ (see Eqs.~\Eq{Qcorr} and \Eq{Qcorri}).

The electrowetting number of case (ii) is given by
\begin{align}
   \eta_\pm(U) 
   & \simeq \frac{\lambda_A}{\lambda_S}\bigg(
            \frac{\epsilon_\text{vac}}{8\gamma_{12}\lambda_S}U^2 
            \pm\frac{\epsilon_\text{vac}(\bar\psi_A-\bar\psi_B)}
                    {2\gamma_{12}\lambda_A}U\bigg)
   \label{eq:ewnimmiscibleii}\\
   & \simeq \frac{\lambda_A}{\lambda_S}\ 
            \frac{\epsilon_\text{vac}}{8\gamma_{12}\lambda_S}U^2, \quad
     \text{for $|U|\gg 4\frac{\lambda_S}{\lambda_A}|\bar\psi_A-\bar\psi_B|$.}
   \notag
\end{align}
This expression bears some resemblance to Eq.~\Eq{phobwall} except of the
typically very small prefactor $\lambda_A/\lambda_S\ll1$ here.

Therefore, electrowetting is also expected to be strongly suppressed for two
immiscible electrolyte solutions with $\epsilon_{F_1}I_{F_1}\not\approx
\epsilon_{F_2}I_{F_2}$, a condition which is typically fulfilled.

For completeness the rather special case (iii) is mentioned, for which the
electrowetting number reads
\begin{align}
   \eta_\pm(U) \simeq 
   \mp\frac{p\epsilon_\text{vac}(\bar\psi_A-\bar\psi_B)}
           {2\gamma_{12}\lambda_S}U.
   \label{eq:ewnimmiscibleiii}
\end{align}


\section{\label{sec:conclusions}Conclusions and Summary}

Electrowetting is studied in the present work by analyzing the capillary rise
of a fluid in the environment of another fluid, where at least one of the two
fluids is an electrolyte solution, for Pellat's setup \cite{Pellat1894, 
Pellat1895} (Fig.~\ref{fig:1}) within the density functional theory of wetting.
Here, the widely ignored possibility of the formation of films of microscopic
thickness on the substrates is taken into account (Fig.~\ref{fig:2}).
Considering the quasi-one-dimensional situation far away from the 
three-phase contact region (Fig.~\ref{fig:3}(a)) allows to transparently derive
the electrowetting equation~\Eq{contactangleeip}.

The derivation shows that electrowetting is a consequence \emph{not} of the
voltage-dependence of the substrate-fluid interfacial tensions, i.e., 
electrowetting is \emph{not} an electrocapillarity effect, \emph{but} of 
the voltage-dependence of the depth of the effective interface potential.
The traditional electrocapillarity approach to electrowetting is shown to
be compromised by the reliance on the incorrect assumption that 
the surface structure of a fluid does not change upon bringing the system
into contact with another fluid phase.

The present analysis of Pellat's setup for electrowetting studies leads to
effectively four length scales corresponding to the Debye lengths in both
fluids, the thickness of the substrates, and the film thicknesses, the latter
being assumed to be approximately equal here, which serve to classify various
relevant experimental situations, e.g., uncoated metal electrodes, 
hydrophilic or hydrophobic dielectric substrates, or fluids comprising
water+oil systems or immiscible electrolyte solutions.
The full dependence of the electrowetting number on these length scales is
derived here (Eq.~\Eq{ewn2}), which can be used for an
actual experimental system.
By considering limiting cases of general interest it is found that for 
uncoated metal electrodes (Eqs.~\Eq{ewnmetalphil} and \Eq{ewnmetalphob})
and wetting of hydrophobic dielectric substrates by water in an oil environment 
(Eq.~\Eq{phobwall}) the electrowetting number within the present
density functional approach agrees with that within the electrocapillarity
picture as well as with numerous experimental studies.
However, a significantly reduced tendency of electrowetting is predicted
here as compared to the predictions within the electrocapillarity approach
for electrowetting on hydrophilic dielectric substrates 
(Eq.~\Eq{philwall}) or situations with both fluids being immiscible
electrolyte solutions (Eqs.~\Eq{ewnimmisciblei}--\Eq{ewnimmiscibleiii}).
Due to a lack of experimental data, verification of these predictions
is an open issue.

One can conclude that it is a matter of fortune that the traditional
electrocapillarity approach to electrowetting (Eqs.~\Eq{eweec2} and
\Eq{CSalpha}) leads to an electrowetting equation, which, although derived by
means of physically questionable arguments, turns out to be yet a precise
approximation for many practical cases.
However, the present study highlights conditions for which significant 
deviations from the electrocapillarity picture are expected to be
experimentally detectable.
The density functional theory of electrowetting presented here is suggested
to be considered as an approach to fundamentally understand as well as to
reliably quantify the phenomenon of electrowetting.





\end{document}